\documentclass[12pt,a4paper]{article}
\usepackage{epsfig}
\def\pslash{\rlap{\hspace{0.02cm}/}{p}}

\begin{document}
\title{ $H_{TC}\Pi^0$ and $\Pi^+\Pi^-$ pair productions at the planned
$e^ + e^ -$ colliders in the topcolor-assisted technicolor model}

\author{Xuelei Wang$^{(a,b)}$, Qingpeng Qiao$^{(b)}$, $Qiaoli Zhang^{(b)}$   \\
 {\small a: CCAST (World Laboratory) P.O. BOX 8730. B.J. 100080 P.R. China}\\
 {\small b: College of Physics and Information Engineering,}\\
\small{Henan Normal University, Xinxiang  453007. P.R.China}
\thanks{This work is supported by the National Natural Science
Foundation of China, the Excellent Youth Foundation of Henan
Scientific Committee, the Henan Innovation Project for University
Prominent Research Talents.}
\thanks{E-mail:wangxuelei@sina.com}}

\maketitle

\begin{abstract}
The top-pions($\Pi^{0,\pm}$) and the top-Higgs($H_{TC}$) are the
typical particles predicted by the topcolor-assisted technicolor
(TC2) model and the observation of these particles can be regarded
as the direct evidence of the TC2 model. In this paper, we study
two pair production processes of these new particles, i.e., $e^ +
e^ - \to H_{TC}\Pi^0$ and $e^+e^-\to \Pi^+\Pi^-$. The results show
that the production rates are at the level of serval fb for $e^ +
e^ - \to H_{TC}\Pi^0$ and tens fb for $e^+e^-\to \Pi^+\Pi^-$.
These processes can produce the adequate distinct multi-jet final
states and the SM background can be efficiently reduced. So, with
the high luminosity at the planned linear colliders, the top-pions
and top-Higgs can be observable via these two pair production
processes without double.
\end{abstract}
\vspace{1.0cm} \noindent
 {\bf PACS number(s)}: 12.60Nz, 14.80.Mz,.15.LK, 14.65.Ha
\newpage

\section{Introduction}

Because the Higgs boson predicted in the standard model(SM) has
not been found experimentally and the mechanism of the electroweak
symmetry breaking(EWSB) is still unknown, one believes that the SM
is only an effective theory of the underlying theory at the TeV
energy scale. Such underlying theory  beyond the SM is called the
new physics model. There are several candidates of new physics
models(such as: the minimal supersymmetric standard model(MSSM),
the technicolor (TC)\cite{TC} model, the top sea-saw
 model \cite{seesaw1,seesaw2} and extra dimensions model\cite{extra}). One of the most
important tasks of the future high energy colliders is to search
for the Higgs in the SM or to test the new physics models,
furthermore, to reveal the mystery of electroweak symmetry
breaking mechanism. In all case, it is likely that the LHC running
in 2007 will see first signals of the mechanism at work.  However,
the complementary information from the linear colliders is needed
to understand the underlying theory.  Several laboratories in the
world have been working on the linear $e^+e^-$ collider projects
with an energy from several hundred GeV up to several TeV and the
yearly expected luminosity being more like $100 fb^{-1}$ or a
little more. These are NLC(USA)\cite{NLC}, JLC(Japan)\cite{JLC},
TESLA(Europe)\cite{TESLA}. The running of these high energy and
luminosity linear colliders will open an unique window for us to
understand the basic theory of particle physics.

As a dynamical breaking theory, the technicolor(TC) model was
introduced by Weinberg and Susskind in the late 1970's\cite{TC},
it can avoid the shortcomings of triviality and
unnaturalness\cite{TC,TC-1} arising from the elementary Higgs
field. Among the various TC models, the topcolor-assisted
technicolor model(TC2)\cite{Hill,Lane,Georgi} is a more realistic
one. One of the most general predictions of the TC2 model is the
existence of three isospin-triplet pseudo Goldstone bosons called
top-pion($\Pi^\pm$,$\Pi^0$) and an isospin-singlet boson called
top-Higgs($H_{TC}$). These bosons can be regarded as the typical
feature of the TC2 model. Thus, study of the production mechanism
of these typical particles can provide us the useful information
to probe them at the future high-energy colliders. The discovery
of these new particles can be regarded as the direct evidence to
test the TC2 model.

The new particles predicted by the TC2 model can be probed
directly via its decay modes. The decay modes of the neutral
top-pion are the tree-level decay processes $\Pi^0 \to
t\bar{t}$(if this is kinetically allowed), $\Pi^0 \to t\bar{c}$,
$\Pi^0 \to b\bar{b}$ and the processes $\Pi^0 \to \gamma \gamma
,gg,\gamma Z$ through an internal top quark loop. The branching
ratios of these possible decay modes have been calculated in
detail\cite{yue}. The results show that the neutral top-pion
almost decays to $t\bar{t}$ if the mass of neutral top-pion
$M_{\Pi}$ is larger than $2m_t$. If $m_t+m_c<M_{\Pi}<2m_t$, the
flavor-changing mode $\Pi^0 \to t\bar{c}$ will become the dominant
decay mode and the branching ratio can be over $60\%$. Such
flavor-changing mode can play an important role in the search of
the neutral top-pion due to the clean background. For the
top-Higgs, the difference case is that there exist tree-level
decay modes $ZZ,W^+W^-$. For the charged top-pions, its decay
modes have been studied in reference \cite{wenna}. The main modes
are the tree-level $\Pi^+ \to t\bar{b}$ and $\Pi^+ \to c\bar{b}$.
The alternative way to probe the top-pions and top-Higgs is to
study the production mechanism of these particles. Recently, we
have systematically studied
 some single top-pion production processes in the
high energy $e^+e^-$\cite{ee}, $\gamma \gamma$\cite{rr} and
$e\gamma $\cite{er} collisions. For the neutral top-pion, we found
that the cross section of $e\gamma \to e\Pi^0$ is over 10 fb, and
the cross sections of the production modes $\Pi^0Z$ in $e^ + e^ -
$ collision and $\Pi^0t\bar{t},\Pi^0t\bar{c}$ in $\gamma\gamma$
collision are at the level of a few fb. With the high luminosity
at the planned linear colliders, the signals should be enough for
us to observe the neutral top-pion. Specially, with the large
cross section, the process $e\gamma \to e\Pi^0$ can not only give
us more information about the TC2 model but also provide an unique
way to distinguish the TC2 model from the SM and other new physics
models. At $e^+e^-$ colliders, the main single charged top-pion
production processes are $e^+e^-\to t\bar{b}\Pi^-$ and $e^+e^-\to
W^+\Pi^-$ with the cross section at order of 10 fb for $e^+e^-\to
t\bar{b}\Pi^-$ and
 a few fb for $e^+e^-\to W^+\Pi^-$. Because the SM
predicts the existence of one neutral Higgs boson, the distinguish
of Higgs-like neutral top-pion with the Higgs in the SM need more
precise measurement, but any observation of charged Higgs or
Higgs-like particles will mean the signal of new physics.
Therefore, the probing of charged top-pions is more important to
test the TC2 model. Besides the single top-pion production
mechanism discussed above, the pair productions $H_{TC}\Pi^0$ and
$\Pi^+\Pi^-$ can occur at the tree-level in the $e^+e^-$
collision. These pair productions might provide more typical
information of the TC2 model via the multi-jets. For the heavy
top-pions and top-Higgs, such processes can only occur at the
planned high energy colliders.
 In this paper, we
study these pair production modes at the planned $e^ + e^ -$
colliders. We find these two processes can provide  enough number
of distinct signals and the SM background can be efficiently
reduced. These processes can provide a feasible way to test the
TC2 model.

As it is known, the $\Pi^{0}$, $H_{TC}$ and $\Pi^{\pm}$ look like
the heavy $A^0$, $H^0$ and $H^{\pm}$ in two-Higgs doublet
model(2HDM), respectively.\footnote{Many production and decay
channels of $A^0$, $H^0$ and $H^{\pm}$ are very similar to that of
$\Pi^{0}$, $H_{TC}$ and $\Pi^{\pm}$. For example, like $H_{TC}$,
$H^0$ can also be produced via $e^+e^-\to A^0H^0, ZH^0,
\bar{\nu}\nu H^0, e^+e^-H^0$ and decay to $f\bar{f}, ZZ,W^+W^-$.}
To distinguish the scalar in the TC2 model from the Higgs sectors
in the 2HDM, we need to point out what differnt features exist
between them. model and the Higgs sectors in the 2HDM. Firstly,
the Yukawa couplings of top-pions and top-Higgs to quarks are
larger than these of Higgs sectors which can make the signals of
the TC2 model become more significant. Secondly, in the TC2 model,
the couplings of top-pions to the three family fermions are
non-universal and the top-pions and top-Higgs have the large
Yukawa couplings to the third family, without the GIM, there exist
large flavor-changing couplings, such as: $\Pi^0t\bar{c}$.
Therefore, the flavor-changing processes in the TC2 model is very
important for us to search for the scalars in the TC2 model and
distinguish such scalars from the Higgs sectors. Thirdly, we find
that there are some difference in the production between the
scalars in the TC2 model and Higgs sectors. For example, there
exist flavor-changing production processes $e^+e^-\to
t\bar{c}\Pi^0(H_{TC}), \gamma\gamma \to t\bar{c}\Pi^0(H_{TC}),
pp\to t\bar{c}\Pi^0(H_{TC})$ which can produce enough signals to
observe the $\Pi^0(H_{TC})$. Such flavor-changing production
processes for Higgs sectors can be ignored due to the GIM. On the
other hand, due to the large Yukawa couplings between $\Pi^0$ and
top quark, the loop-level $e^+e^-\to Z(\gamma)\Pi^0$ become more
important than the similar process $e^+e^-\to Z(\gamma)A^0$.
Finally, there are also some difference in the decay modes between
them. If $t\bar{t}$ mode is forbidden, with large branching ratio,
$t\bar{c}$ become the main decay mode for $\Pi^0$ and $H_{TC}$.
But such decay mode does not exist for Higgs sectors. It should
also be noted that $b\bar{b}$ is an important decay mode for Higgs
sectors with large $tan \beta$, but for the scalars in the TC2
model, such decay mode can be ignored due to the small coupling of
scalars to $b\bar{b}$. The above discusion can help us to
distinguish the scalars in the TC2 model from the Higgs sectors in
2HDM.

 The rest parts of this paper is organized as
follows. In section II, we firstly present a brief review of the
TC2 model and then calculate the cross sections of the processes
$e^ + e^ - \to H_{TC}\Pi^0$ and $e^+e^-\to \Pi^+\Pi^-$. The
numerical results are also discussed in section II. The
conclusions are given in Section III.

\section{The production cross sections of $e^ + e^ - \to
H_{TC}\Pi^0$ and $e^+e^-\to \Pi^+\Pi^-$}
\subsection{A brief review of the TC2 model}
The large top quark mass is suggestive of new dynamics associated
with electroweak symmetry breaking. In order to solve the heavy
top quark and the EWSB problems, the topcolor model was proposed
\cite{topcolor}. In  the topcolor model, the top quark
participates in a new strong interaction(topcolor) which is
spontaneously broken at some high energy scale $\Lambda_t$. The
strong dynamics leads to the formation of a condensate
$<t\bar{t}>$ and gives rise to a large dynamical mass for top
quark. If this top condensate is to be an adequate source of
electroweak symmetry breaking, and at the same time, gives a
reasonable top quark mass, the scale $\Lambda_t$ must be very
high($10^{15}$ GeV). i.e., the topcolor model suffers unnatural
problem. Another dynamical breaking theory, technicolor model, can
solve the problem related to Higgs field in the SM, but such
theory has been unable to provide a natural and plausible
understanding of why the top quark mass is so large. The TC2
model\cite{Hill}, incorporating the best features of the
technicolor and topcolor, offers a new insight into possible
mechanism of EWSB and the origin of heavy top quark mass. At EWSB
scale, this model predicts two groups of scalars corresponding to
the technicolor condensates and topcolor condensates,
respectively. Either of them can be arranged into a $SU(2)$
doublet\cite{Rainwater}, and their roles in TC2 model are quite
analogous to the Higgs fields in the model proposed in
Ref.\cite{special} which is a special two-Higgs-doublet model in
essence. Explicitly speaking, the doublet $\Phi_{TC}$ which
corresponds to the topcolor condensates plays a minor role in EWSB
and only couples to the third generation quarks, its main task is
to generate the large top quark mass. While the doublet
$\Phi_{ETC}$ which corresponds to the technicolor condensates is
mainly responsible for EWSB and light fermion masses, it also
contributes a small portion of top quark mass. The vacuum
expectation value(vev) of the top quark pair condensate $f_{\pi}$
can  be given by the Pagel-Stokar formula. For condensation around
the EWSB scale of 1 TeV, $f_{\pi}$ should be near 60 GeV. Once
$f_{\pi}$ is fixed, the vev of the technifermion condensates,
$v_T$, is uniquely determined by the EWSB requirement
$f_{\pi}^2+v_T^2=v^2\simeq (246 GeV)^2$. For $f_{\pi}=60$ GeV, we
must have $v_T=239$ GeV. We linearize the theory and rearrange the
pions in two orthogonal linear combinations to form the
longitudinal degrees of freedom of the weak gauge bosons and a
triplet of top-pions, $\Pi^{0,\pm}$, which become physical degrees
of freedom. The top-pions are analogous to the neutral CP-odd and
charged Higgs scalars of two-Higgs doublet model(2HDM). The theory
loosely predicts top-pions to lie in the mass range of 200 GeV.
Besides physical top-pions, there are another two CP-even Higgs
modes, labeled $H_{TC}$ and $H_{ETC}$, are known as the
"top-Higgs" boson and the "techni-Higgs" bosons, respectively.
Their masses can be estimated in the Nambu-Jona-lasinio(NJL) model
in the large $N_c$ approximation. For the top-Higgs boson this is
found to be on the order of $M_H\simeq 2m_t$; for the techni-Higgs
boson it is much higher. However, it only serves as a rough guide.
From the kinetic terms of the effective TC2 Lagrangian in
linearized form\cite{Rainwater}
\begin{equation}
L_{kin}=(D_{\mu}\Phi_{TC})^{\dag}(D^{\mu}\Phi_{TC})
+(D_{\mu}\Phi_{ETC})^{\dag}(D^{\mu}\Phi_{ETC}),
\end{equation}
we know there exist tree-level couplings $Z^{\mu}H_{TC}\Pi^0$,
$Z^{\mu}\Pi^+\Pi^-$ and $A^{\mu}\Pi^+\Pi^-$ which can induce the
tree-level production processes $e^+e^-\to H_{TC}\Pi^0$ and
$e^+e^-\to \Pi^+\Pi^-$\footnote{There should be another production
$e^+e^-\to H_{ETC}\Pi^0$, but the cross section of such process is
strongly depressed by  the factor$\frac{f_{\pi}}{v}$ and heavy
$H_{ETC}$. On the other hand, techni-Higgs is not the typical
particle of the TC2 model, so, we do not study this process in
this paper.}. These couplings can be written as:
\begin{equation}
Z^{\mu}H_{TC}\Pi^0: \hspace{0.1cm}
-i\frac{g}{2c_w}\frac{v_T}{v}(p_{\mu}^{H}-p_{\mu}^0),
\hspace{0.4cm}Z^{\mu}\Pi^-\Pi^+:  \hspace{0.1cm}
\frac{g}{c_w}(1-2s^2_w)(p_{\mu}^{-}-p_{\mu}^+),
\hspace{0.4cm}A^{\mu}\Pi^-\Pi^+:\hspace{0.1cm}e(p_{\mu}^{-}-p_{\mu}^+).
\end{equation}
Where, $c_w = \cos \theta _w(\theta _w$ is the Weinberg angle).

\subsection{The processes $e^+e^-\to H_{TC}\Pi^0$}
With the coupling $Z^{\mu}H_{TC}\Pi^0$, the process $e^+e^-\to
H_{TC}\Pi^0$ can be induced at tree-level via $Z^0$ exchanging.
The Feynman diagram of the process is shown in Fig.1(a). The
production amplitude of the process can be written directly:
\begin{eqnarray}
 M_{H_{TC}\Pi^0}=\frac{2i\pi\alpha_e v_T}{s_w^2c_w^2v}\bar{v}_{e^+}(p_1
)(\pslash_3-\pslash_4)(-\frac{1}{2}L+s^2_w)u_{e^-}(p_2)G(p_1+p_2,M_Z).
\end{eqnarray}
Where, $G(p,M)=\frac{1}{p^2 - M^2 }$ denotes the propagator of the
particle, $p_3$ and $p_4 $ denote the momenta of outcoming
top-Higgs and neutral top-pion, $L=\frac{1-\gamma_5}{2}$. With the
above production amplitude, we can directly obtain the production
cross section of the process $e^ + e^ - \to H_{TC}\Pi^0$.

To obtain numerical results of the cross section, we fixed the
input parameters as: $M_Z=91.187$ GeV, $s^2_w=0.23$, $f_{\pi}=60$
GeV, the mass of top-Higgs $M_H=350$ GeV($\approx2m_t$). The
electromagnetic fine-structure constant $\alpha_{e}$ at a certain
energy scale is calculated from the simple QED one-loop evolution
with the boundary value $\alpha_{e}=\frac{1}{137.04}$. Although
the theory predicts top-pions to lie in the mass range of 200 GeV,
this can be only regarded as a rough guide. In order to give a
general prediction, we expand the mass range to 150-400 GeV. To
show the influence of the center of mass energy($\sqrt{s}$) on the
cross section, we take $\sqrt{s}=800,1600$
GeV,respectively($\sqrt{s}$=500 is too small to produce
$H_{TC}\Pi^0$). The numerical results of the cross section are
shown in Fig 2.

The plots show that the cross section decreases with $M_{\Pi}$ due
to the phase space depression and falls more sharply for
$\sqrt{s}=800$ GeV. The cross section is not sensitive to
$M_{\Pi}$ when $\sqrt{s}=1600$ GeV. The change of the cross
section with $\sqrt{s}$ is not monotonous because the influence of
$\sqrt{s}$ on the phase space and Z-propagator is inverse. In
general, the production rate is at the level of a few fb. We find
that $\sqrt{s}=800$ GeV is an ideal energy to probe light top-pion
. For the light top-pion, the production rate can be near 10 fb in
the case of $\sqrt{s}=800$. With yearly expected luminosity about
$100 fb^{-1}$, there are $10^3-10^4$ $H_{TC}\Pi^0$ signals can be
produced after several running of $e^+e^-$ colliders.

As has been discussed, the possible decay modes of $\Pi^0$ are the
tree-level decay modes: $t\bar {t}$(if $\Pi^0
> 2m_t$), $t\bar{c}$, $b\bar {b}$ and loop-level decay modes: $gg$,
$\gamma \gamma$, $Z\gamma $. As it is known, the couplings of
top-pion to the three generation fermions are non-universal and
therefore do not possess a Glashow-Iliopoulos-Maiani(GIM)
mechanism, this non-universal feature results in a large
flavor-changing coupling $\Pi^0t\bar {c}$. So, in the case of
light $\Pi^0$, the main decay mode should be $\Pi^0\to t\bar{c}$.
In the SM, the cross section of the process with $t\bar {c}$
production is strongly depressed by GIM mechanism. Therefore,
$\Pi^0\to t\bar{c}$ might provide the typical signals of the TC2
model. With $M_H\simeq 2m_t$, the main decay modes of top-Higgs
should be tree-level modes $t\bar{c}$,$Z^0Z^0$ and
$W^+W^-$($H_{TC}\to t\bar{t}$ might be forbidden or its decay
width is very small when $M_H$ is near $2m_t$). With above
discussion, we know that the most interesting signals of the
$H_{TC}\Pi^0$  production should be the four quark flavor-changing
jets :$t\bar{t}t\bar{c}, t\bar{c}t\bar{c}$ which can provide
distinct signals to identify the $H_{TC}\Pi^0$ production with the
large production rate and clean SM background\footnote{In the SM,
the production rates of $t\bar{t}t\bar{c},t\bar{c}t\bar{c}$ are
very small due to the GIM mechanism.}. On the other hand,
$e^+e^-\to H_{TC}\Pi^0$ can produce a large number of
$t\bar{t}c\bar{c}$ which can also be produced via $e^+e^-\to
A^0H^0$ in the 2HDM, but the different pole structure make it
possible for us to distinguish them, the pole structure can be
easily detected at the $e^+e^-$ colliders. This means that
reconstructing the neutral top-pion from $t\bar{c}$ will be
important both as a means for measuring the mass of top-pion and
also as a means for identifying the neutral scalars via
flavor-changing $t\bar{c}$ jet.

In the 2HDM, there exists a similar process $e^+e^-\to A^0H^0$. To
distinguish the scalars in the TC2 model from the Higgs in the
2HDM via $e^+e^-\to H_{TC}\Pi^0$, we should compare the cross
sections of $e^+e^-\to H_{TC}\Pi^0$ and $e^+e^-\to A^0H^0$. The
process $e^+e^-\to A^0H^0$ has been studied in
reference\cite{2HDM}. We find the the cross section behavior of
$e^+e^-\to A^0H^0$ is similar to that of $e^+e^-\to H_{TC}\Pi^0$.
The cross sections values of such two processes are not
significantly different for the same parameters. For example,
$\sigma(e^+e^-\to H_{TC}\Pi^0)=4.08, 3.95 fb$ for
$\sqrt{s}=800,1600$ GeV and $M_{\Pi}=300$ GeV, $\sigma(e^+e^-\to
A^0H^0)=5,3.9 fb$ for $\sqrt{s}=800,1600$ GeV and $M_{A}=300$ GeV.
 We should distinguish the
neutral $H_{TC},\Pi^0$ depending on their different feature of
decay modes and pole structure. $\Pi^0(H_{TC})\to t\bar{c}$ can
provide the typical signals of the TC2 model. For neutral Higgs
boson, $b\bar{b}, \tau^+\tau^-$ are important decay modes,
specially for large $tan\beta$, but the decay modes $b\bar{b}$ can
be ignored and $\tau^+ \tau^-$ does not exist for $H_{TC},\Pi^0$.
On the other hand, the neutral Higgs boson can decay to
supersymmetric particles and there are not similar decay modes for
$H_{TC},\Pi^0$.

\subsection{The process $e^+e^-\to \Pi^+\Pi^-$}
The charged top-pion pair $\Pi^+\Pi^-$ can be produced in $e^+e^-$
annihilation via virtual $\gamma$ or $Z^0$ exchanging as shown in
Figs.1(b). Such process is very important in searching for the
charged top-pions because it can produce the distinct signals of
the charged top-pions. The production amplitudes are:
\begin{eqnarray}
 M^{\gamma}_{\Pi^+\Pi^-}=4\pi\alpha_e\bar {v}_{e^ + } (p_1
)(\pslash_4-\pslash_3)u_{e^-}(p_2)G(p_1+p_2,0)
 \end{eqnarray}
\begin{eqnarray}
 M^{Z}_{\Pi^+\Pi^-}=-\frac{4\pi\alpha_e}{s^2_wc_w^2}(1-2s^2_w)\bar {v}_{e^ + } (p_1
)(\pslash_4-\pslash_3)(s^2_w-\frac{L}{2})u_{e^-}(p_2)G(p_1+p_2,M_Z)
 \end{eqnarray}

We ignore the electroweak contribution to the masses of charged
top-pions and take $M_{\Pi^+}=M_{\Pi^-}=M_{\Pi}$. The cross
section of $e^+e^-\to \Pi^+\Pi^-$ is shown in Fig.3.

It is shown that the behavior of the cross section plots of
$e^+e^-\to \Pi^+\Pi^-$ versus $M_{\Pi}$ is similar to that of
$e^+e^-\to H_{TC}\Pi^0$. For $\sqrt{s}=500$ GeV, the cross section
falls sharply to a very small rate with $M_{\Pi}$ increasing. So,
the energy 500 GeV is not suitable to search for the heavy charged
top-pion pair. In the most case, the cross section is at the order
of tens fb which is significantly larger than that of $e^+e^-\to
H_{TC}\Pi^0$(The process $e^+e^-\to \Pi^+\Pi^-$ includes an extra
$\gamma-$propagator contribution). Such production rate
corresponds about $10^3$ $\Pi^+\Pi^-$ pairs with the luminosity
$100^{-1} fb$. This abundant production allows to enforce tight
requirements on the event pre-selection and the mass
reconstruction. The most promising decay modes to search for the
charged top-pions are $\Pi^+\to t\bar{b}$ and the flavor-changing
mode $\Pi^+\to c\bar{b}$. In the case of $\Pi^+\to t\bar{b}$, the
signals of the charged top-pion pair production is
$t\bar{t}b\bar{b}$. In order to efficiently distinguish the
signals from the underlying backgrounds and to measure the
top-pion mass, it is important to obtain a clean charged top-pion
signal in the mass distribution of the multi-jet final states. To
identify the production mode $t\bar{t}b\bar{b}$, we insist on 8
jets or 1 lepton plus 6 jets(in particular, fewer than 10 visible
lepton/jets so as to discriminate from the 4t final states) and
possibly require that one W and the associated t be reconstructed.
In particular, since final states contain at least four b jets, in
order to eliminate any residual QCD background, we need one or two
b-tags without incurring significant penalty. Such b-tagging
should have efficiency of $60\%$ or better. For the light charged
top-pions, the branching ratio of $\Pi^+\to c\bar{b}$ can be
comparative to that of $\Pi^+\to t\bar{b}$. In this case,
$\Pi^+\to c\bar{b}$ is also an important mode which induces the
signals $c\bar{b}\bar{c}b$. Although $\Pi^+\to c\bar{b}$ is a
flavor-changing decay mode, $c\bar{b}\bar{c}b$ production is not
the flavor-changing process. Therefore, the SM background can not
be ignored. The major irreducible background should come from
$e^+e^-\to Z^0Z^0\to c\bar{c}b\bar{b}$. The mistagging of b-quark
and s-quark will make the $e^+e^-\to W^+W^-$ become important
which significantly enhances the background. So, the efficient b
tagging and mass reconstruction of the charged top-pion is very
necessary to reduce the background.

To compare the cross section of $e^+e^-\to \Pi^+\Pi^-$ with that
of similar process $e^+e^-\to H^+H^-$, we find that the former is
significantly larger than the latter \footnote{For example,
$\sigma(e^+e^-\to \Pi^+\Pi^-)$=21.51, 15.09 fb with
$\sqrt{s}=800,1600$ GeV and $M_{\Pi}$=300 GeV, $\sigma(e^+e^-\to
H^+H^-)$=10.02,8.1 fb with $\sqrt{s}$=800,1600 GeV and $M_{H}$=300
GeV.} which provide some useful information to distinguish the
charged top-pions from charged Higgs. The $t\bar{b}$ is the main
decay mode for both charged top-pions and charged Higgs, such mode
is not suitable to distinguish these particles. To obtain the
identified signals of the charged top-pions, we should probe
charged top-pions via the flavor-changing decay mode $\Pi^+\to
c\bar{b}$. $\tau\nu_{\tau}$ can also provide the identified
signals of charged Higgs which does not exist for the charged
top-pions.

 The production rates of the processes $e^+e^-\to H_{TC}\Pi^0$
and $e^+e^-\to \Pi^+\Pi^-$ are studied in this section. In order
to give a believable prediction of the yearly event rate, we also
need to consider the overall efficiency factor for detector
coverage and for experimentally isolating and detecting these
modes. Such efficiency can be reasonable estimated as $40\%$. On
the other hand, the mass reconstruction and b-tagging will also
reduce the efficiency to detect the signals. Considering these
efficiencies, we can safely estimate that there should be abundant
signals can be identified with the high luminosity at the planned
linear colliders, and furthermore, the masses of these new
particles might be measured with high accuracy.

\section{The conclusions}

In the framework of the TC2 model, $H_{TC}\Pi^0$ and $\Pi^+\Pi^-$
pair productions at the planned $e^ + e^ -$ colliders are studied
in this paper. We find that the production rates are at the level
of a few fb for $H_{TC}\Pi^0 $ production and tens fb for
$\Pi^+\Pi^-$ production. These pair productions can produce
multi-jet final states and the SM background can be efficiently
reduced. We conclude that the top-pions and top-Higgs predicted by
the TC2 model should be experimentally observable via these
processes at the planned colliders with high luminosity.
Therefore, $H_{TC}\Pi^0$ and $\Pi^+\Pi^-$ pair productions at $e^+
e^-$ colliders are very promising production mechanism of the
top-pions and top-Higgs.

\newpage
\newpage
\begin{figure}[ht]
\begin{center}
\begin{picture}(250,200)(0,0)
\put(-80,-70){\epsfxsize 140 mm \epsfbox{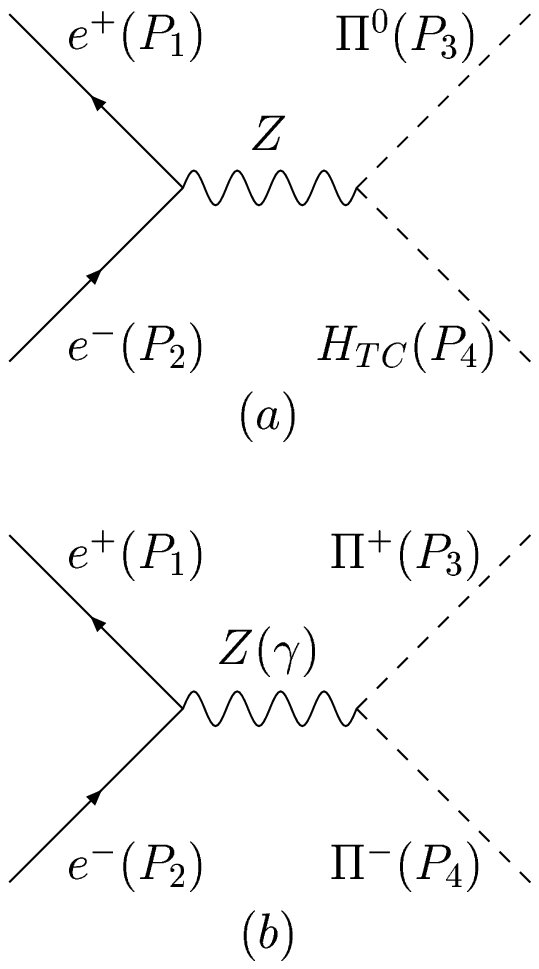}}
\put(-10,-70){Fig.1.The Feynman diagrams of the processes
 $e^ + e^ - \to H_{TC}\Pi^0$ and $e^ + e^ - \to \Pi^+\Pi^-$.}
\end{picture}
\end{center}
\end{figure}

\begin{figure}[hb]
\begin{center}
\begin{picture}(250,200)(0,0)
\put(-80,-180){\epsfxsize 140 mm \epsfbox{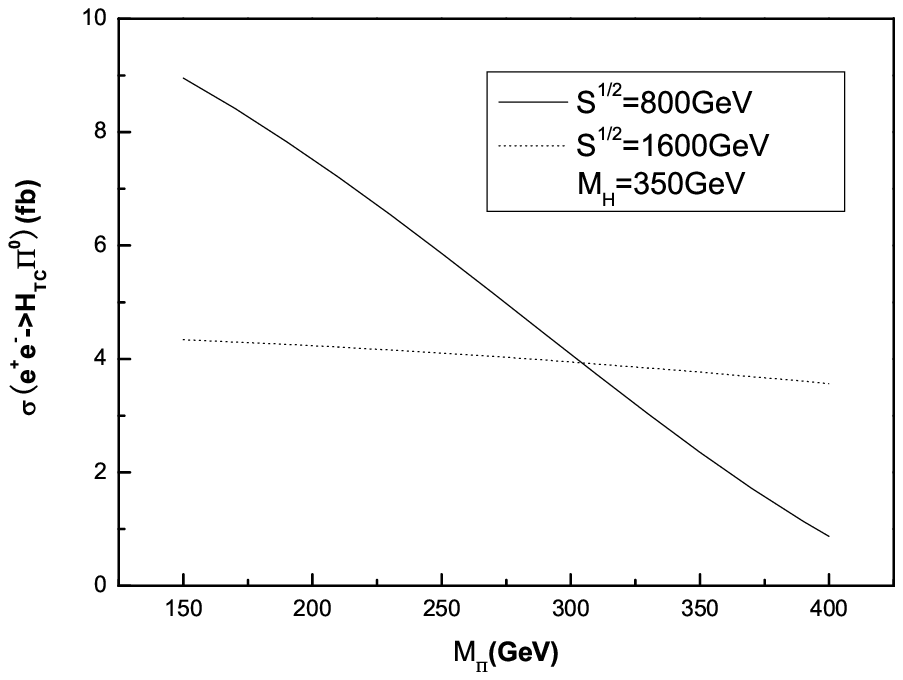}}
\put(-10,-180){Fig.2. The cross section of $e^ + e^ - \to
H_{TC}\Pi^0$ versus top-pion mass $M_{\Pi}$(150-400 GeV) for
$M_H=350$ GeV and $\sqrt{s}$=800 GeV(solid line),1600 GeV(dot
line), respectively.}
\end{picture}
\end{center}
\end{figure}


\newpage
\begin{figure}[ht]
\begin{center}
\begin{picture}(250,200)(0,0)
\put(-80,-70){\epsfxsize 140 mm \epsfbox{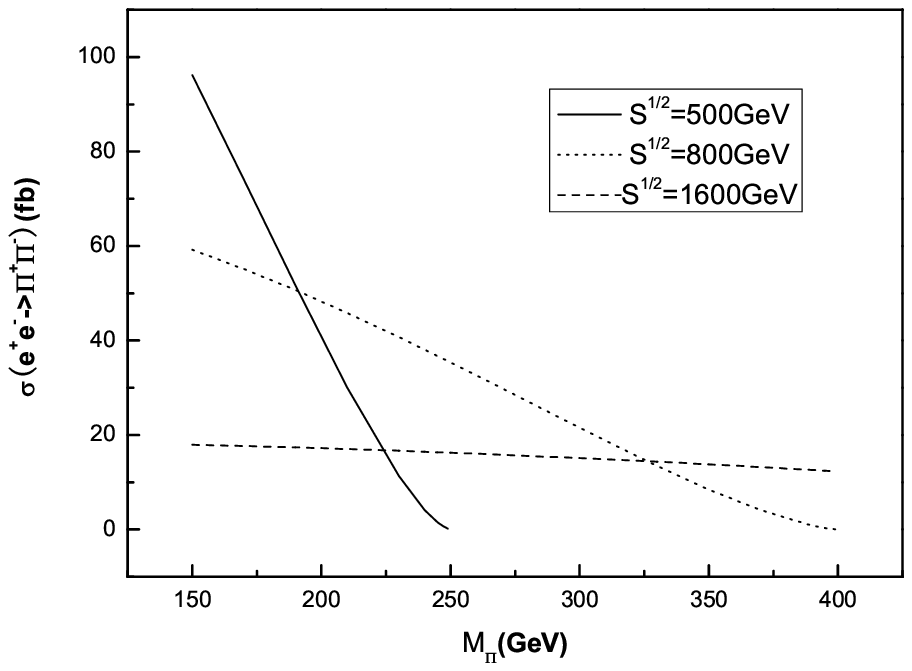}}
\put(-10,-70){Fig.3. The cross section of $e^ + e^ - \to
\Pi^+\Pi^-$ versus top-pion mass $M_{\Pi}$(150-400 GeV) with
$\sqrt{s}$=500 GeV(solid line), 800 GeV(dot line),1600 GeV(dash
line), respectively. }
\end{picture}
\end{center}
\end{figure}


\end{document}